\pdfoutput=1
\documentclass[]{collinear}
\usepackage[T1]{fontenc}
\usepackage[utf8]{inputenc}

\usepackage{lmodern}
\usepackage{microtype}
\usepackage{amsmath,amssymb}
\usepackage{booktabs,graphicx,xcolor}
\usepackage{tabularx}
\title{CWE-bench: Refocusing Agentic Security from Exploitation to Defense}
\author[*]{Gonzalo~Gonzalez-Pumariega}
\author[*]{Ajay~Jagannath}
\author{Adit~Jain}
\author{Guram~Gogia}
\author{\newline Soumyadeep~Bakshi}
\author{Nazneen~Rajani}
\contribution[*]{Equal contribution}

\usepackage{amsfonts}
\usepackage{hyperref}
\usepackage{url}
\usepackage{booktabs}
\usepackage{xspace}
\usepackage{amsmath}
\usepackage{tabularx}
\usepackage{multirow}
\usepackage{enumitem}
\usepackage{titlesec}
\usepackage{listings}
\usepackage{xcolor}
\usepackage{float}
\usepackage[dvipsnames]{xcolor}
\usepackage{fvextra}
\usepackage{float}
\usepackage[T1]{fontenc}
\usepackage[utf8]{inputenc} 
\usepackage{tcolorbox}
\tcbuselibrary{breakable}
\usepackage{fvextra,times}
\usepackage{newunicodechar}
\titlespacing{\paragraph}{0pt}{0.5em}{0.5em}
\abstract{
Cybersecurity evaluations of AI agents have focused largely on offense, yet a working exploit does not show that an agent can make software safe.
An attacker needs to exploit only one vulnerability; a defender must find and fix all of them.
We introduce CWE-bench, a benchmark of 250 repository audit-and-repair tasks spanning 185 CWE types and eight languages, grouped into ten weakness classes. We release 100 tasks publicly and keep 150 held out.
Agents are given a repository to audit and a description of its threat model, but are not told the weakness, where it occurs, or how many vulnerable sites there are.
Each patch must block the exploit while preserving original functionality, as checked by a deterministic programmatic verifier. Verifiers were curated with panels of agentic judges and reviewed by hand, so that they reject incomplete repairs without penalizing correct ones.
The best agent reaches 67\% pass@1 and 79\% pass@4.
We further conduct analyses showing that (i) agents with similar pass@1 differ widely in retry gains and cost, and (ii) no single agent is best across languages or weakness classes. As the offensive capabilities of agents improve, CWE-bench provides a broad and rigorous measure of whether their defensive capabilities keep pace.

}
\correspondence{\url{research@collinear.ai}}
\coldata[Website]{\url{https://collinear.ai}}
\date{\today}

\begin{document}

\maketitle

\section{Introduction}
\label{sec:introduction}

Can an agent make a vulnerable codebase safe to use? Most cybersecurity evaluations of AI agents measure offense: solving capture-the-flag challenges, exploiting applications, or reproducing vulnerabilities~\citep{cybench2025,cvebench2025,cybergym2026}. Defense has received far less attention, even though it is the harder problem. An attacker needs to exploit only one vulnerability; a defender must find and fix all of them without breaking the software. Consider a file server with a path-traversal flaw. An attacker needs just one handler that builds a file path from user input to read files outside the shared directory. A defender must find and fix every such handler, and cannot simply disable downloads, since that removes the service the server exists to provide.

A growing number of benchmarks evaluate defensive agents, but each simplifies the problem in a different way. Many build tasks directly on public Common Vulnerabilities and Exposures (CVEs), whose fixes may already be in training data, or on LLM-generated code bases that fall outside the distribution of real software~\citep{patcheval2025,cvefactory2026,vul4py2026,secprobe2026}. Others give the agent a vulnerability description, proof-of-concept input, or location hints, reducing discovery to repair of a known flaw~\citep{patcheval2025,patchbench2026,vul4py2026}. The rest cover only a narrow slice of weaknesses, such as memory-safety bugs in C/C++~\citep{cybergyme2e2026,patchbench2026,zerodaybench2026}. A real defender faces none of these simplifications: it must find unknown vulnerabilities across the full range of weaknesses in real code.

We introduce \textbf{CWE-bench}\footnote{Tasks, leaderboard, and results: \url{https://cwe-bench.com}.}, a benchmark of 250 repository auditing and repair tasks that addresses each of these gaps. Every task is built on a real open-source repository with injected vulnerabilities. We release 100 tasks publicly, with their verifiers, reference patches, and agent rollouts, so that anyone can reproduce our results and evaluate new agents; the remaining 150 are held out, so their injected vulnerabilities have no published fix. Agents receive only the repository and an audit instruction describing the system and its threat model; they are not told the weakness, where it occurs, or how many vulnerable sites there are. Tasks span 185 Common Weakness Enumeration (CWE) types and eight programming languages, which we group into ten weakness classes. A patch passes only if it blocks the exploits while preserving the original functionality, as checked by a deterministic programmatic verifier. To make the verifiers both hard and fair, we curated them with panels of agentic judges and reviewed every task by hand. Table~\ref{tab:intro-coverage} compares CWE-bench with recent defensive benchmarks.

The best agent, Claude Opus 5.5, reaches 67\% pass@1 and 79\% pass@4, and the four leading agents lie within three points of pass@1. Beyond these aggregate scores, agents with similar pass@1 differ widely in how much they gain from retries and in cost per attempt, and their ordering changes across languages and weakness classes, with no single agent leading in either.

The paper makes three contributions:
\begin{enumerate}
\item \textbf{CWE-bench}, a benchmark of 250 auditing and repair tasks on real repositories, covering 185 CWEs across ten weakness classes with no diagnosis given to the agent. We release 100 tasks and 4{,}400 agent rollouts, and keep 150 tasks held out.
\item \textbf{A deterministic grading procedure} that checks both exploit blocking and preserved functionality, with verifiers curated by panels of agentic judges and human review.
\item \textbf{An evaluation of eleven frontier agents}, showing that similar pass@1 hides large differences in retry gains, task coverage, and cost, and that no single agent is best across languages or weakness classes.
\end{enumerate}
\begin{table}[!tb]
\centering
\footnotesize
\setlength{\tabcolsep}{3pt}
\renewcommand{\arraystretch}{1.1}
\begin{tabularx}{\linewidth}{@{}l r r c r l l >{\raggedright\arraybackslash}X@{}}
\toprule
Benchmark & Tasks & CWEs & OWASP 2025 & Lang. & Codebase & Vuln.\ source & Information given to agent \\
\midrule
PatchEval~\citep{patcheval2025} & 230 & 39 & 7/10 & 3 & Real & Public CVEs & Vulnerability description \\
CyberGym-E2E$^\dagger$~\citep{cybergyme2e2026} & 920 & NR & \textit{1/10} & 2 & Real & Public OSS-Fuzz bugs & None \\
LiveCVEBench~\citep{cvefactory2026} & 190 & 74 & 9/10 & 14 & Real & Public CVEs & Bug report \\
ZeroDayBench$^\dagger$~\citep{zerodaybench2026} & 22 & \textit{15} & \textit{4/10} & \textit{6} & Real & Injected & Varies (5 levels) \\
PatchBench~\citep{patchbench2026} & 213 & 16 & 1/10 & 2 & Real & Injected & PoC and sanitizer report \\
Vul4Py~\citep{vul4py2026} & 100 & 60 & 7/10 & 1 & Real & Public CVEs & CVE and file/line hints \\
SecProbe$^\dagger$~\citep{secprobe2026} & 353 & 151 & \textit{5/10} & 6 & LLM-generated & Injected & Software specification \\
\midrule
\textbf{CWE-bench} & \textbf{250} & \textbf{185} & \textbf{10/10} & \textbf{8} & \textbf{Real} & \textbf{Injected} & \textbf{Threat model} \\
\bottomrule
\end{tabularx}
\caption{Comparison of recent vulnerability repair benchmarks. \emph{OWASP 2025} counts the OWASP Top 10:2025 categories covered under each category's official CWE mapping. \emph{Codebase} indicates whether tasks are built on real open-source repositories or LLM-generated ones. \emph{Injected} denotes vulnerabilities inserted into a codebase, such as real CVEs transplanted into new repositories. C/C++ and JavaScript/TypeScript count as separate languages. PatchEval counts its 230-task sandboxed subset. CWE-bench counts its 100 public and 150 held-out tasks. NR: not reported. $^\dagger$Some values (in italics) are derived by us: from source-CVE records (ZeroDayBench), sanitizer crash types (CyberGym-E2E), or the 20 most frequent CWEs reported (SecProbe, a lower bound).}
\label{tab:intro-coverage}
\end{table}
\section{Related Work}
\label{sec:related-work}

\paragraph{Offensive evaluation.}
Offensive success is self-certifying: a captured flag, a triggered crash, or a working exploit is proof on its own. This has let offensive evaluation scale quickly, from capture-the-flag challenges~\citep{cybench2025} to exploiting real web applications~\citep{cvebench2025}, reproducing real vulnerabilities~\citep{cybergym2026}, achieving consequential attacks~\citep{exploitgym2026}, and grading intermediate exploitation steps~\citep{exploitbench2026}, with multi-stage penetration tests now separating tooling gaps from planning failures~\citep{deng2026pentesting}. Defensive success is not self-certifying. Blocking the known exploit is necessary but not sufficient, since a patch can stop one payload while leaving variants open, or disable the feature entirely. Offensive benchmarks therefore show whether agents can expose weaknesses, but not whether they can remove them.

\paragraph{Defensive auditing and repair.}
Defensive benchmarks work around the grading problem through shortcuts that limit what they measure. First, many build tasks on public CVEs and grade against the developer's own fix and tests~\citep{secbench2025,patcheval2025,cvefactory2026,vul4py2026}. Those fixes are published and may appear in training data, so a correct patch can reflect recall rather than investigation. Generating the repository along with its tests avoids contamination~\citep{secprobe2026}, but produces code outside the distribution of real software. Second, many give the agent a vulnerability description, proof-of-concept input, or location hints~\citep{patcheval2025,patchbench2026,vul4py2026,cvefactory2026}. The agent then only repairs a known flaw and skips auditing, a large part of real defensive work. Third, many cover only a narrow range of weaknesses~\citep{cybergyme2e2026,patchbench2026,zerodaybench2026}. Several are limited to memory-safety bugs in C/C++, where sanitizers give an automatic check, leaving access-control, authentication, and logic flaws untested; these reach only one to four of the ten OWASP Top 10:2025 categories (Table~\ref{tab:intro-coverage}). CWE-bench addresses all three: real repositories with injected vulnerabilities, no diagnosis given to the agent, and coverage of 185 CWEs spanning all ten OWASP Top 10:2025 categories. Because the held-out tasks have no published fix, CWE-bench can also release part of its evaluation set for reproducibility while keeping a held-out portion that cannot be recalled.

\paragraph{Evaluation validity and integrity.}
Even with executable checks, defensive grading can go wrong in three ways. First, test-passing patches can still be incorrect~\citep{qi2015plausibility}. Prior work strengthens validation with fuzzing and differential checks, maintainer-added tests, or semantic checks~\citep{autopatchbench2025,pvbench2026,patchbench2026}, and finds errors in proof-of-concept grading itself~\citep{secbenchpro2026}. Second, LLM judges are not a reliable substitute: they can agree with one another while disagreeing with execution-based checks~\citep{deininger2026fixing}. Third, agents can game the evaluation by retrieving published fixes or exploiting weaknesses in the evaluation harness~\citep{kouremetis2026cheats,benchjack2026,benchshield2026}. CWE-bench combines these lessons: a deterministic programmatic verifier provides the score, panels of agentic judges and human reviewers shaped those verifiers during curation, every verifier is checked to reject the published upstream fix, and agents can reach only their model API during evaluation.

\section{Benchmark Design}
\label{sec:benchmark}

\subsection{Problem formulation}
\label{sec:formulation}

\begin{figure}[t]
\centering
\includegraphics[width=\linewidth]{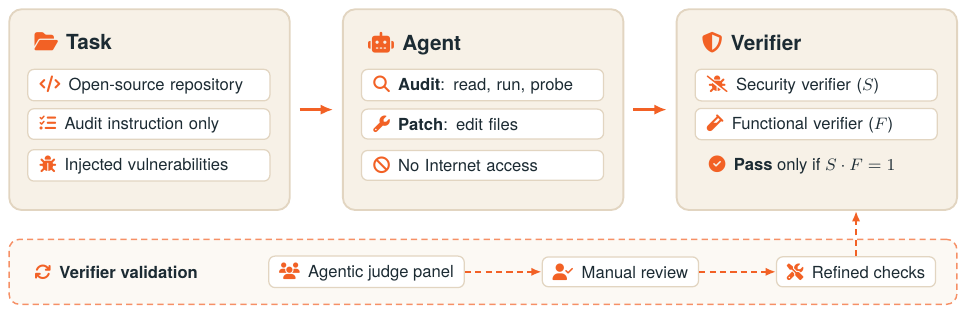}
\caption{CWE-bench evaluation. Each task provides an open-source repository with injected vulnerabilities and an audit instruction that does not name them. The agent audits and patches the repository without Internet access. A patch passes only if both the security verifier ($S$) and the functional verifier ($F$) succeed. During curation, the verifiers are reviewed by panels of agentic judges and by hand to refine the checks.}
\label{fig:workflow}
\end{figure}

We formulate each task as a Markov decision process $\mathcal{M} = \langle \mathcal{S}, \mathcal{A}, \mathcal{T}, R \rangle$. A state $s_t \in \mathcal{S}$ consists of the repository, the audit instruction, and the interaction history up to step $t$, where the instruction does not name the vulnerabilities or their number. Actions $a_t \in \mathcal{A}$ are the operations available through the agent's harness, such as inspecting code, running commands, and editing files. The transition function $\mathcal{T}: \mathcal{S} \times \mathcal{A} \rightarrow \mathcal{S}$ applies an action's effect on the repository and appends its output to the history; actions that attempt to access the Internet are blocked and leave the repository unchanged. The agent acts according to a policy $\pi(a_t \mid s_t)$. When the agent submits or reaches the one-hour time limit at step $T$, the reward $R: \mathcal{S} \rightarrow \{0, 1\}$ is computed on the final state, so a timed-out rollout is graded on whatever repairs it has made:
\begin{equation}
R(s_T) = S(s_T) \cdot F(s_T),
\label{eq:reward}
\end{equation}
where $S(s_T) = 1$ if the exploits for the target vulnerabilities are blocked, and $F(s_T) = 1$ if the repository's existing functionality tests still pass. A rollout passes only if both checks succeed. Both checks are deterministic, so a given final state always receives the same reward.

\begin{figure}[t]
 \centering
 \includegraphics[width=\linewidth]{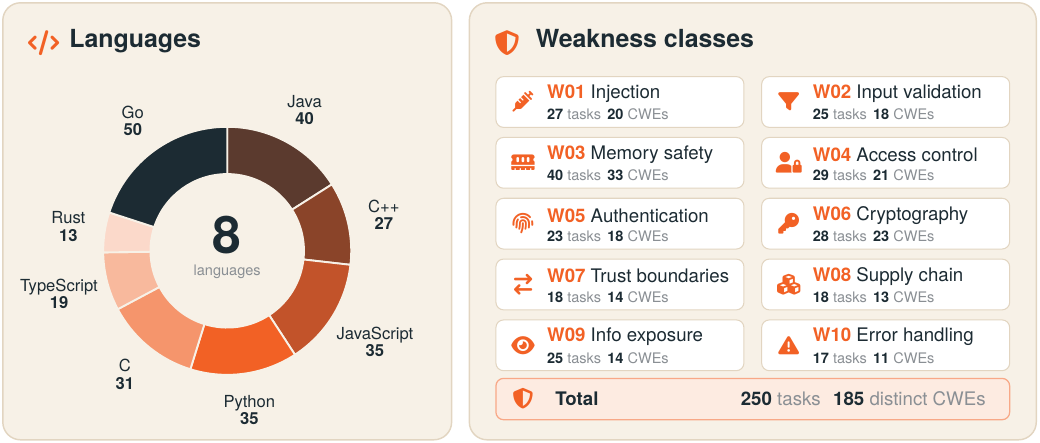}
 \caption{Composition of the 250-task evaluation set (100 public, 150 held-out). \textbf{Left:} tasks per programming language. \textbf{Right:} tasks and distinct CWEs per weakness class (Appendix~\ref{app:task-coverage}).}
 \label{fig:coverage}
\end{figure}

\subsection{Task coverage and construction}

The evaluation set contains 250 tasks spanning 185 CWE types and eight programming languages, from injection and access control to memory-safety errors (use after free, out-of-bounds writes, type confusion), certificate revocation, and LLM prompt injection. Many of these CWEs sit at the edges of the OWASP Top 10:2025 categories or outside them, so we group every CWE into ten weakness classes of our own, in the spirit of OWASP but finer where the set requires it (Appendix~\ref{app:task-coverage}); the set still covers all ten OWASP Top 10:2025 categories. Figure~\ref{fig:coverage} breaks the set down by language and by weakness class; each class contains 17 to 40 tasks.

Each task starts from an open-source repository with a real CVE, into which we inject new vulnerabilities. For each task, we write exploits that trigger the injected vulnerabilities and functionality tests that exercise the repository's intended behavior; these form the checks $S$ and $F$ in Equation~\ref{eq:reward}. A task is accepted only if a reference patch passes both checks and the unmodified repository fails.

\paragraph{Public and held-out sets.} We split the tasks into a public set of 100 tasks, spanning 91 CWEs, and a held-out set of 150 tasks, spanning 126 CWEs. The public set is released with its environments, verifiers, reference patches, and the 4{,}400 agent rollouts behind our results, so that anyone can inspect a task, reproduce our scores, and evaluate new agents (Appendix~\ref{app:release}). The held-out set is kept private: its injected vulnerabilities have no published fix and cannot appear in training data, so it continues to measure progress on tasks no agent has seen. We report results on the full 250-task set and show the two subsets for illustration.

\subsection{Quality checks}
\label{sec:verifier-validation}

A verifier that is too lenient rewards incomplete patches, and one that is too strict penalizes correct patches that differ from the reference fix. We curated the verifiers with panels of agentic judges (Appendix~\ref{app:judge-protocol}). Each judge works in a separate copy of a patched repository, where it can inspect code, build the project, and run the verifier, and assesses the patch on five criteria: security, functionality, robustness to attack variants, grading integrity, and build soundness. We manually reviewed rollouts where the panel and the verifier disagreed (Appendix~\ref{app:human-review}). These reviews led us to add probes that catch incomplete repairs, to replace checks that assumed the reference patch's structure with checks of behavior, and to correct judge rubrics where the judges erred. The panels shape the verifiers only; the reported score is the programmatic verdict alone.

In addition, every task was checked against six questions and reviewed by hand (Table~\ref{tab:task-checks}). The checks ensure that a task reads as a realistic request, that its graded behavior is fair to expect from the instruction, and that its verifier rewards a real repair rather than a memorized one.

\begin{table}[!tb]
\centering
\small
\begin{tabularx}{\linewidth}{lX}
\toprule
Check & Requirement \\
\midrule
Realistic instruction & The instruction reads like a real request, not an evaluation setting. \\
Inferable requirements & Every graded behavior can be inferred from the instruction. \\
Behavior, not structure & The verifier checks for correct behavior, not for the structure of one particular fix. \\
Reference and baseline & The unmodified repository fails and the reference patch passes. \\
Preserved functionality & The original functionality still works after a correct fix. \\
No memorized fixes & The verifier rejects memorized solutions, such as the known upstream patch. \\
\bottomrule
\end{tabularx}
\caption{Checks applied to every task.}
\label{tab:task-checks}
\end{table}

\section{Experiments}
\label{sec:protocol}

\subsection{Experimental Setup}

\paragraph{Agents.} We evaluate eleven frontier models, each run in a coding-agent harness: Claude Code, Codex, Antigravity, or OpenCode (Table~\ref{tab:release-results}). Scores therefore reflect the model and its harness together. Each agent uses high reasoning effort and makes four independent attempts per task, with a one-hour limit per attempt.

\paragraph{Execution.} Each attempt runs in an isolated container where the agent can reach only its model API. Attempts in which the model API declines the request, for example when a safety filter blocks the task, count as failures.

\paragraph{Metrics.} Each attempt is scored by the programmatic verifier (Equation~\ref{eq:reward}). We report pass@1, the mean fraction of passing attempts per task, and pass@4, the fraction of tasks with at least one passing attempt~\citep{chen2021code}, along with the mean API cost per rollout. The leaderboard is computed on all 250 tasks; we also report the public and held-out sets separately for illustration.

\begin{table}[!tb]
\centering
\small
\setlength{\tabcolsep}{4pt}
\begin{tabular}{llrrrrrrr}
\toprule
 & & \multicolumn{2}{c}{All 250 tasks} & \multicolumn{2}{c}{Public (100)} & \multicolumn{2}{c}{Held-out (150)} & Mean cost \\
\cmidrule(lr){3-4}\cmidrule(lr){5-6}\cmidrule(lr){7-8}
System & Harness & Pass@1 & Pass@4 & Pass@1 & Pass@4 & Pass@1 & Pass@4 & (USD) \\
\midrule
Claude Opus 5.5 & Claude Code & \textbf{67.1} & \textbf{79.2} & 67.2 & 78.0 & \textbf{67.0} & \textbf{80.0} & 1.85 \\
Muse Wonder & OpenCode & 66.1 & 78.8 & 64.8 & 79.0 & \textbf{67.0} & 78.7 & 2.73 \\
Gemini 4 Argon & Antigravity & 65.0 & 77.2 & \textbf{67.8} & \textbf{81.0} & 63.2 & 74.7 & 3.70 \\
GPT-6 Astra & Codex & 64.2 & 72.8 & 63.8 & 70.0 & 64.5 & 74.7 & 3.87 \\
DeepSeek V4.1 Flash & OpenCode & 57.6 & 73.6 & 57.0 & 72.0 & 58.0 & 74.7 & 0.04 \\
Grok 4.7 & OpenCode & 54.7 & 71.2 & 55.5 & 74.0 & 54.2 & 69.3 & 2.97 \\
Hy4 Preview & OpenCode & 47.3 & 64.4 & 52.0 & 68.0 & 44.2 & 62.0 & 0.43 \\
Mistral Large 4 & OpenCode & 45.6 & 60.4 & 46.2 & 59.0 & 45.2 & 61.3 & 0.37 \\
Qwen 3.8 Max & OpenCode & 45.4 & 64.8 & 45.0 & 62.0 & 45.7 & 66.7 & 2.06 \\
Kimi K3.1 Pro & OpenCode & 45.3 & 64.0 & 41.5 & 63.0 & 47.8 & 64.7 & 0.96 \\
GLM 5.3 & OpenCode & 42.0 & 63.6 & 44.0 & 65.0 & 40.7 & 62.7 & 0.58 \\
\bottomrule
\end{tabular}
\caption{Pass@1 and pass@4 (\%) on all 250 tasks, which form the leaderboard, and on the public and held-out sets, with mean API cost per rollout over all 250 tasks. Agents are ordered by pass@1 on all tasks; the best value in each column is in bold.}
\label{tab:release-results}
\label{tab:reference-results}
\end{table}

\subsection{Main Results}
\label{sec:release-results}

Table~\ref{tab:release-results} reports results for all eleven agents. Claude Opus 5.5 leads with 67.1\% pass@1 and 79.2\% pass@4, followed by Muse Wonder (66.1\%), Gemini 4 Argon (65.0\%), and GPT-6 Astra (64.2\%). The public and held-out sets rank the agents similarly (Spearman $\rho = 0.82$ across the eleven agents), with a mean gap of 2.7 points in pass@1 between the two sets. The leading agent differs between them, however: Gemini 4 Argon leads on the public set and Claude Opus 5.5 and Muse Wonder tie on the held-out set. Of the 250 tasks, 230 are solved by at least one agent and 20 by none.

\subsection{How do repeated attempts and different agents expand task coverage?}
\label{sec:coverage}

\begin{figure}[!tb]
\centering
\includegraphics[width=\linewidth]{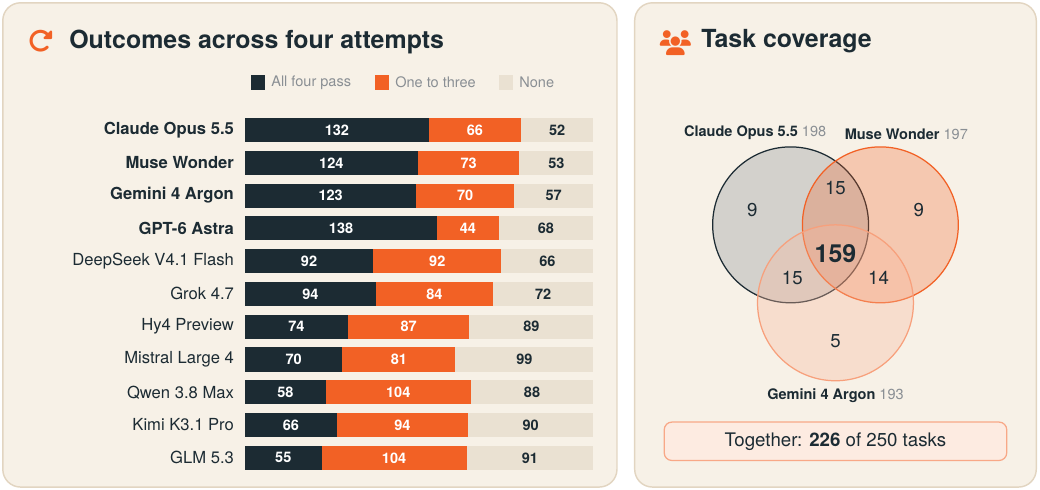}
\caption{Outcomes across four attempts and task coverage. \textbf{Left:} for each agent, the number of tasks passed on all four attempts, on one to three, and on none; the four leading agents are in bold. \textbf{Right:} tasks passed at least once by each of the three leading agents, and their overlap.}
\label{fig:release-consistency}
\end{figure}

The four leading agents lie within three points of pass@1, yet range from 72.8\% to 79.2\% pass@4 (Table~\ref{tab:release-results}). The difference lies in how consistently each agent repairs the tasks it can solve (Figure~\ref{fig:release-consistency}, left). GPT-6 Astra passes all four attempts on 138 tasks, more than any other agent, but has mixed outcomes on only 44, so additional attempts add little and it has the lowest pass@4 of the four. Claude Opus 5.5 passes all four attempts on fewer tasks (132) but has mixed outcomes on 66, which retries convert into passes. The same pattern holds further down the table: DeepSeek V4.1 Flash has mixed outcomes on 92 tasks and reaches a higher pass@4 than GPT-6 Astra (73.6\% against 72.8\%) despite a pass@1 6.6 points lower.

The leading agents also solve different tasks (Figure~\ref{fig:release-consistency}, right). Claude Opus 5.5, Muse Wonder, and Gemini 4 Argon pass at least one attempt on 198, 197, and 193 tasks, respectively, but only 159 are common to all three, and together they pass 226 of the 250. Claude Opus 5.5 passes 33 tasks that GPT-6 Astra never does, while GPT-6 Astra passes 17 that Claude Opus 5.5 never does.

\subsection{How does pass rate scale with inference cost?}
\label{sec:cost}

\begin{figure}[!tb]
\centering
\includegraphics[width=\linewidth]{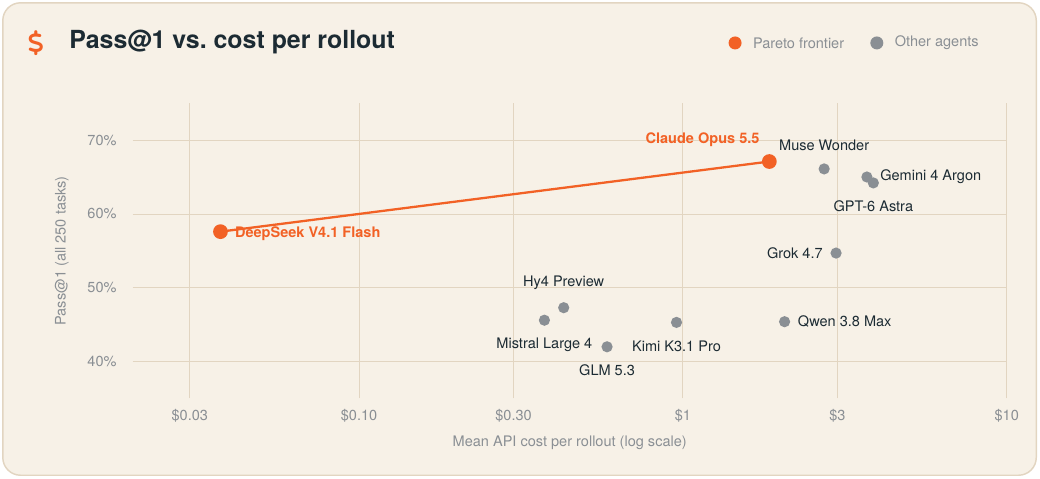}
\caption{Pass@1 against mean API cost per rollout, using the values in Table~\ref{tab:release-results}. Orange points mark the Pareto frontier: agents for which no agent of equal or lower cost scores higher.}
\label{fig:release-cost}
\end{figure}

Claude Opus 5.5 reaches the highest pass@1 (67.1\%) at \$1.85 per rollout, while Muse Wonder, Gemini 4 Argon, and GPT-6 Astra score lower at \$2.73, \$3.70, and \$3.87 (Figure~\ref{fig:release-cost}). DeepSeek V4.1 Flash reaches 57.6\% at \$0.04. These two agents form the Pareto frontier. Along it, the step from DeepSeek V4.1 Flash to Claude Opus 5.5 costs about 50$\times$ more per rollout for 9.5 more points of pass@1, or 1.2$\times$ the pass rate.

\subsection{How does performance vary across languages and weakness classes?}
\label{sec:subgroups}

\begin{figure}[!htb]
\centering
\includegraphics[width=\linewidth]{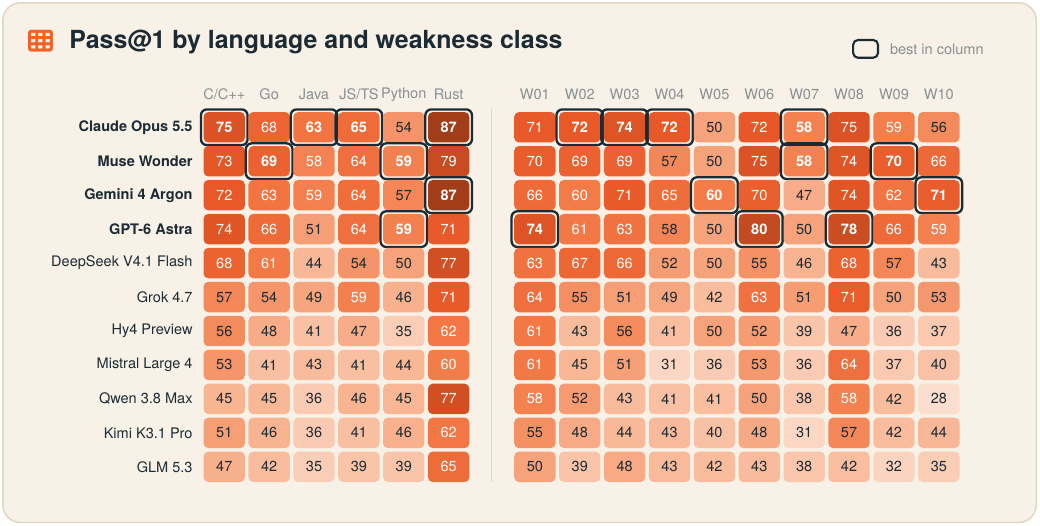}
\caption{Pass@1 by language (left) and weakness class (right). C/C++ and JS/TS are grouped. Outlined cells mark the best score in each column.}
\label{fig:release-subgroups}
\end{figure}

The leading agents trade places across both languages and weakness classes (Figure~\ref{fig:release-subgroups}). Claude Opus 5.5 leads on C/C++ (75.0\%), JS/TS (65.3\%), and Java (62.5\%), Muse Wonder on Go (69.0\%), and Muse Wonder and GPT-6 Astra tie on Python (58.6\%). Across weakness classes, the leader changes more often: Claude Opus 5.5 leads three of the ten classes outright, GPT-6 Astra three, Gemini 4 Argon two, and Muse Wonder one, and Claude Opus 5.5 and Muse Wonder tie on the remaining one, trust boundaries and request handling (W07). W07 and authentication and sessions (W05) are the hardest classes: the best agent reaches 58.3\% and 59.8\%, and the mean agent 44.7\% and 46.5\%. Supply chain and data integrity (W08) and cryptography and secrets (W06) are among the easiest, with GPT-6 Astra reaching 77.8\% and 79.5\%.

\section{Discussion}
\label{sec:discussion}

CWE-bench asks agents to find and repair vulnerabilities in real repositories without being told the weakness or where it occurs, across 185 CWEs in ten weakness classes. Across eleven agents, the leaders lie within three points of pass@1 but differ in retries, coverage, and cost, and the leading agent changes across languages and weakness classes. We draw three takeaways from these results.

\paragraph{Automated vulnerability repair is becoming practical.}
The leading agents audit and repair most tasks end to end, without being told the weakness or where it occurs, and reach 79\% pass@4 with retries. This performance is no longer limited to the most expensive models: Claude Opus 5.5 leads at a lower cost per rollout than each of the next three agents (\$1.85 against \$2.73 to \$3.87), and DeepSeek V4.1 Flash reaches 58\% pass@1 at \$0.04 per rollout (Section~\ref{sec:cost}), which makes repair cheap enough to run routinely rather than only in periodic audits. Agents also solve different tasks: the three leading agents together cover 226 of the 250 tasks, against 198 for the best single agent (Section~\ref{sec:coverage}). Systems that combine several agents, rather than relying on one, are a natural direction for automated defense.

\paragraph{Narrow benchmarks give a misleading picture of security capability.}
The leading agent changes from one weakness class to the next, and the four leading agents each lead at least one class (Section~\ref{sec:subgroups}). A benchmark that covers a few weakness classes, as several defensive benchmarks do~\citep{cybergyme2e2026,patchbench2026,zerodaybench2026}, can therefore rank agents differently than one that covers many, and a single aggregate score hides where each agent is weak. CWE-bench covers 185 CWEs, but this is still a fraction of the 956 active weaknesses in CWE (Appendix~\ref{app:task-coverage}). Judging how ready agents are for defensive work requires a broader map of vulnerability classes, with results reported per class rather than as one number.

\paragraph{Verifiers must be hard but fair, and open to inspection.}
A benchmark is only as good as what its verifier accepts. A verifier that is too lenient rewards incomplete patches, such as one that blocks the known exploit but leaves variants open, and overstates how ready an agent is to secure code. A verifier that is too strict penalizes correct patches that remove the vulnerability and preserve behavior but differ from the expected fix. Both errors mislead anyone choosing an agent, and the same holds for any team that uses automated checks to accept agent-written patches. CWE-bench addresses this in two ways. During curation, panels of agentic judges and human reviewers shaped each verifier toward checks of behavior that reject incomplete repairs and memorized fixes (Section~\ref{sec:verifier-validation}). After curation, we release 100 tasks with their verifiers, reference patches, and rollouts, so that anyone can inspect what a verifier accepts and rejects and report a task they believe is wrong, while the 150 held-out tasks continue to measure progress on tasks no agent has seen.

\section{Conclusion}
\label{sec:conclusion}

Security evaluations of AI agents have largely measured how well agents attack software. As agents are deployed to audit and maintain code, the more consequential question is whether they make software safer. CWE-bench measures this directly: agents must find and repair unlabeled vulnerabilities in real repositories, and a patch counts only if it blocks the exploit and preserves the program's behavior. Our results show that agents differ in how they respond to retries, which tasks they cover, what they cost, and which weakness classes they handle, differences that a single pass rate hides. We release 100 tasks and their rollouts so that others can reproduce these results and evaluate new agents, and keep 150 tasks held out to track progress on tasks no agent has seen. As the offensive capabilities of agents improve, their defensive capabilities must keep pace, and the only way to know whether they do is to measure defense rigorously across the full breadth of vulnerabilities that real software contains.

\bibliographystyle{unsrtnat}
\bibliography{references}

@inproceedings{cybench2025,
 author = {Zhang, Andy K and Perry, Neil and Dulepet, Riya and Ji, Joey and Menders, Celeste and Lin, Justin and Jones, Eliot and Hussein, Gashon and Liu, Samantha and Jasper, Donovan and Peetathawatchai, Pura and Glenn, Ari and Sivashankar, Vikram and Zamoshchin, Daniel and Glikbarg, Leo and Askaryar, Derek and Yang, Haoxiang and Zhang, Aolin and Alluri, Rishi and Tran, Nathan and Sangpisit, Rinnara and Oseleononmen, Kenny and Boneh, Dan  and Ho, Daniel and Liang, Percy},
 booktitle = {International Conference on Learning Representations},
 editor = {Y. Yue and A. Garg and N. Peng and F. Sha and R. Yu},
 pages = {25094--25243},
 title = {Cybench: A Framework for Evaluating Cybersecurity Capabilities and Risks of Language Models},
 url = {https://proceedings.iclr.cc/paper_files/paper/2025/file/3e9412a9c1d93810ef3ef7825115016b-Paper-Conference.pdf},
 volume = {2025},
 year = {2025}
}

@InProceedings{cvebench2025,
  title = 	 {{CVE}-Bench: A Benchmark for {AI} Agents’ Ability to Exploit Real-World Web Application Vulnerabilities},
  author =       {Zhu, Yuxuan and Kellermann, Antony and Bowman, Dylan and Li, Philip and Gupta, Akul and Danda, Adarsh and Fang, Richard and Jensen, Conner and Ihli, Eric and Benn, Jason and Geronimo, Jet and Dhir, Avi and Rao, Sudhit and Yu, Kaicheng and Stone, Twm and Kang, Daniel},
  booktitle = 	 {Proceedings of the 42nd International Conference on Machine Learning},
  pages = 	 {79850--79867},
  year = 	 {2025},
  editor = 	 {Singh, Aarti and Fazel, Maryam and Hsu, Daniel and Lacoste-Julien, Simon and Berkenkamp, Felix and Maharaj, Tegan and Wagstaff, Kiri and Zhu, Jerry},
  volume = 	 {267},
  series = 	 {Proceedings of Machine Learning Research},
  month = 	 {13--19 Jul},
  publisher =    {PMLR},
  url = 	 {https://proceedings.mlr.press/v267/zhu25i.html}
}

@inproceedings{cybergym2026,
 author = {WANG, ZHUN and Shi, Tianneng and He, Jingxuan and Cai, Matthew and Zhang, Jialin and Song, Dawn},
 booktitle = {International Conference on Learning Representations},
 editor = {C. Vondrick and B. Hariharan and C. Raffel and L. Pinto and D. Yang and A. Faust},
 pages = {123341--123386},
 title = {CyberGym: Evaluating AI Agents\textquotesingle  Real-World Cybersecurity Capabilities at Scale},
 url = {https://proceedings.iclr.cc/paper_files/paper/2026/file/c876a2935a90aff21874e14c07dc3e33-Paper-Conference.pdf},
 volume = {2026},
 year = {2026}
}

@misc{exploitgym2026,
      title={ExploitGym: Can AI Agents Turn Security Vulnerabilities into Real Attacks?}, 
      author={Zhun Wang and Nico Schiller and Hongwei Li and Srijiith Sesha Narayana and Milad Nasr and Nicholas Carlini and Xiangyu Qi and Eric Wallace and Elie Bursztein and Luca Invernizzi and Kurt Thomas and Yan Shoshitaishvili and Wenbo Guo and Jingxuan He and Thorsten Holz and Dawn Song},
      year={2026},
      eprint={2605.11086},
      archivePrefix={arXiv},
      primaryClass={cs.CR},
      url={https://arxiv.org/abs/2605.11086}, 
}

@misc{exploitbench2026,
      title={ExploitBench: A Capability Ladder Benchmark for LLM Cybersecurity Agents}, 
      author={Seunghyun Lee and David Brumley},
      year={2026},
      eprint={2605.14153},
      archivePrefix={arXiv},
      primaryClass={cs.CR},
      url={https://arxiv.org/abs/2605.14153}, 
}

@misc{deng2026pentesting,
      title={What Makes a Good LLM Agent for Real-world Penetration Testing?}, 
      author={Gelei Deng and Yi Liu and Yuekang Li and Ruozhao Yang and Xiaofei Xie and Jie Zhang and Han Qiu and Tianwei Zhang},
      year={2026},
      eprint={2602.17622},
      archivePrefix={arXiv},
      primaryClass={cs.CR},
      url={https://arxiv.org/abs/2602.17622}, 
}

@inproceedings{secbench2025,
 author = {Lee, Hwiwon and Zhang, Ziqi and Lu, Hanxiao and ZHANG, LINGMING},
 booktitle = {Advances in Neural Information Processing Systems},
 doi = {10.52202/085713-3878},
 editor = {D. Belgrave and C. Zhang and H. Lin and R. Pascanu and P. Koniusz and M. Ghassemi and N. Chen},
 pages = {116342--116378},
 publisher = {Curran Associates, Inc.},
 title = {SEC-bench: Automated Benchmarking of LLM Agents on Real-World Software Security Tasks},
 url = {https://proceedings.neurips.cc/paper_files/paper/2025/file/a9168f1c54e5147027f1e8cf83e1a775-Paper-Conference.pdf},
 volume = {38, Main Conference},
 year = {2025}
}

@misc{patcheval2025,
      title={PATCHEVAL: A New Benchmark for Evaluating LLMs on Patching Real-World Vulnerabilities}, 
      author={Zichao Wei and Jun Zeng and Ming Wen and Zeliang Yu and Kai Cheng and Yiding Zhu and Jingyi Guo and Shiqi Zhou and Le Yin and Xiaodong Su and Zhechao Ma},
      year={2025},
      eprint={2511.11019},
      archivePrefix={arXiv},
      primaryClass={cs.CR},
      url={https://arxiv.org/abs/2511.11019}, 
}

@InProceedings{cvefactory2026,
  title = 	 {{CVE}-Factory: Scaling Expert-Level Agentic Tasks for Code Security Vulnerability},
  author =       {Luo, Xianzhen and Zhang, Jingyuan and Zhou, Shiqi and Huang, Jinyang and Xiao, Chuan and Zhu, Qingfu and Ma, Zhiyuan and Xing, Yue and Yue, Yang and Zeng, Wencong and Che, Wanxiang},
  booktitle = 	 {Proceedings of the 43rd International Conference on Machine Learning},
  pages = 	 {83137--83168},
  year = 	 {2026},
  editor = 	 {Zhang, Tong and Dudik, Miroslav and Jaggi, Martin and Agarwal, Alekh and Li, Sharon and Schuurmans, Dale and Zhu, Jerry and Berkenkamp, Felix and Dong, Hanze and Bietti, Alberto},
  volume = 	 {306},
  series = 	 {Proceedings of Machine Learning Research},
  month = 	 {06--11 Jul},
  publisher =    {PMLR},
  url = 	 {https://proceedings.mlr.press/v306/luo26aj.html}
}

@misc{vul4py2026,
      title={Vul4Py: Benchmarking Automated Vulnerability Repair in Python with Paired Exploit and Functional Oracles}, 
      author={Tan Bui and Ting Zhang and Ferdian Thung and Yunpeng Xiong and Penghao Jiang and Xin Zhou and David Lo},
      year={2026},
      eprint={2608.00692},
      archivePrefix={arXiv},
      primaryClass={cs.SE},
      url={https://arxiv.org/abs/2608.00692}, 
}

@misc{secprobe2026,
      title={SecProbe: Adaptive Evaluation of Coding Agents on Cybersecurity Vulnerabilities}, 
      author={Xiaonan Luo and Yue Huang and Kehan Guo and Ping He and Chuan Zou and Chujie Gao and Lichi Li and Yuchen Ma and Zhangchen Xu and Zichen Chen and Yufei Han and Xiangliang Zhang},
      year={2026},
      eprint={2609.33763},
      archivePrefix={arXiv},
      primaryClass={cs.CR},
      url={https://arxiv.org/abs/2609.33763}, 
}

@misc{patchbench2026,
      title={PatchBench: Evaluating AI Agents for Vulnerability Patching}, 
      author={Chihao Shen and Jiacheng Li and Aastha Mahajan and Jeffery Siyuan Tian and Yonghwi Kwon and Yizheng Chen},
      year={2026},
      eprint={2609.04075},
      archivePrefix={arXiv},
      primaryClass={cs.CR},
      url={https://arxiv.org/abs/2609.04075}, 
}

@InProceedings{cybergyme2e2026,
  title = 	 {{C}yber{G}ym-{E}2{E}: Scalable Real-World Benchmark for {AI} Agents’ End-to-End Cybersecurity Capabilities},
  author =       {Shi, Tianneng and Rheem, Robin and Jiang, Dongwei and Wang, Mona and De La Riega, Francisco and Wang, Zhun and Jiang, Jingzhi and Cheung, Alexander and Tai, Sean and Cha, Jonah and Tu, Jianhong and Han, Gabriel and Wang, Chenguang and He, Jingxuan and Guo, Wenbo and Song, Dawn},
  booktitle = 	 {Proceedings of the 43rd International Conference on Machine Learning},
  pages = 	 {112094--112108},
  year = 	 {2026},
  editor = 	 {Zhang, Tong and Dudik, Miroslav and Jaggi, Martin and Agarwal, Alekh and Li, Sharon and Schuurmans, Dale and Zhu, Jerry and Berkenkamp, Felix and Dong, Hanze and Bietti, Alberto},
  volume = 	 {306},
  series = 	 {Proceedings of Machine Learning Research},
  month = 	 {06--11 Jul},
  publisher =    {PMLR},
  url = 	 {https://proceedings.mlr.press/v306/shi26r.html}
}

@misc{zerodaybench2026,
      title={ZeroDayBench: Evaluating LLM Agents on Unseen Zero-Day Vulnerabilities for Cyberdefense}, 
      author={Nancy Lau and Louis Sloot and Jyoutir Raj and Giuseppe Marco Boscardin and Evan Harris and Dylan Bowman and Mario Brajkovski and Jaideep Chawla and Dan Zhao},
      year={2026},
      eprint={2603.02297},
      archivePrefix={arXiv},
      primaryClass={cs.CR},
      url={https://arxiv.org/abs/2603.02297}, 
}

@inproceedings{qi2015plausibility, series={ISSTA ’15}, title={An analysis of patch plausibility and correctness for generate-and-validate patch generation systems}, url={http://dx.doi.org/10.1145/2771783.2771791}, DOI={10.1145/2771783.2771791}, booktitle={Proceedings of the 2015 International Symposium on Software Testing and Analysis}, publisher={ACM}, author={Qi, Zichao and Long, Fan and Achour, Sara and Rinard, Martin}, year={2015}, month=July, pages={24–36}, collection={ISSTA ’15} }

@misc{autopatchbench2025,
  author = {Byun, TJ and Aschermann, Cornelius and Thng, Kai Yuan and Zhou, Wu and Yang, Yupeng and Deason, Lauren and Saxe, Joshua},
  title = {Introducing {AutoPatchBench}: A Benchmark for {AI}-Powered Security Fixes},
  year = {2025},
  month = apr,
  howpublished = {Engineering at Meta},
  note = {Technical release, April 29, 2025. Accessed September 30, 2026},
  url = {https://engineering.fb.com/2025/04/29/ai-research/autopatchbench-benchmark-ai-powered-security-fixes/}
}

@misc{pvbench2026,
      title={Patch Validation in Automated Vulnerability Repair}, 
      author={Zheng Yu and Wenxuan Shi and Xinqian Sun and Zheyun Feng and Meng Xu and Xinyu Xing},
      year={2026},
      eprint={2603.06858},
      archivePrefix={arXiv},
      primaryClass={cs.SE},
      url={https://arxiv.org/abs/2603.06858}, 
}

@misc{secbenchpro2026,
      title={SEC-bench Pro: Can Language Models Solve Long-Horizon Software Security Tasks?}, 
      author={Hwiwon Lee and Jiawei Liu and Dongjun Kim and Wubing Xia and Ziqi Zhang and Chunqiu Steven Xia and Lingming Zhang},
      year={2026},
      eprint={2605.26548},
      archivePrefix={arXiv},
      primaryClass={cs.CR},
      url={https://arxiv.org/abs/2605.26548}, 
}

@article{deininger2026fixing, title={Fixing, Breaking, or Faking It? An Execution-Calibrated Evaluation of LLM Vulnerability Patching in JavaScript, Python, Go, and Java, and the Limits of LLM-as-Judge}, volume={6}, ISSN={2624-800X}, url={http://dx.doi.org/10.3390/jcp6050153}, DOI={10.3390/jcp6050153}, number={5}, journal={Journal of Cybersecurity and Privacy}, publisher={MDPI AG}, author={Deininger, Patrick and Slany, Wolfgang}, year={2026}, month=Sept, pages={153} }

@misc{kouremetis2026cheats,
      title={Every Model Cheats: Prompt-Level Mitigation of Cheating on Offensive Cyber Tasks}, 
      author={Michael Kouremetis and Ads Dawson and Raja Sekhar Rao Dheekonda and Brian Greunke},
      year={2026},
      eprint={2607.21763},
      archivePrefix={arXiv},
      primaryClass={cs.CR},
      url={https://arxiv.org/abs/2607.21763}, 
}

@misc{benchjack2026,
      title={Do Androids Dream of Breaking the Game? Systematically Auditing AI Agent Benchmarks with BenchJack}, 
      author={Hao Wang and Hanchen Li and Qiuyang Mang and Alvin Cheung and Koushik Sen and Dawn Song},
      year={2026},
      eprint={2605.12673},
      archivePrefix={arXiv},
      primaryClass={cs.AI},
      url={https://arxiv.org/abs/2605.12673}, 
}

@misc{benchshield2026,
      title={BenchShield: Formal Model-Backed Instrumentation for Reward Integrity in LLM-Agent Evaluation Infrastructure}, 
      author={Shenghan Zheng and Zonglin Di and Yimin Liu and Kyoung Whan Choe and Jiankai Sun and Heguang Lin and Penghao Jiang and Yifeng He and Xiao Cheng and Jicheng Wang and Wenbo Chen and Alex Yates and Yinzhe Zhao and Bingran You and Yuan Gao and Ayush Munot and Shubham Gaur and Zhe Ye and Hao Wang and Xiangyi Li and Dawn Song and Christophe Hauser},
      year={2026},
      eprint={2609.11028},
      archivePrefix={arXiv},
      primaryClass={cs.CR},
      url={https://arxiv.org/abs/2609.11028}, 
}

@misc{chen2021code,
      title={Evaluating Large Language Models Trained on Code}, 
      author={Mark Chen and Jerry Tworek and Heewoo Jun and Qiming Yuan and Henrique Ponde de Oliveira Pinto and Jared Kaplan and Harri Edwards and Yuri Burda and Nicholas Joseph and Greg Brockman and Alex Ray and Raul Puri and Gretchen Krueger and Michael Petrov and Heidy Khlaaf and Girish Sastry and Pamela Mishkin and Brooke Chan and Scott Gray and Nick Ryder and Mikhail Pavlov and Alethea Power and Lukasz Kaiser and Mohammad Bavarian and Clemens Winter and Philippe Tillet and Felipe Petroski Such and Dave Cummings and Matthias Plappert and Fotios Chantzis and Elizabeth Barnes and Ariel Herbert-Voss and William Hebgen Guss and Alex Nichol and Alex Paino and Nikolas Tezak and Jie Tang and Igor Babuschkin and Suchir Balaji and Shantanu Jain and William Saunders and Christopher Hesse and Andrew N. Carr and Jan Leike and Josh Achiam and Vedant Misra and Evan Morikawa and Alec Radford and Matthew Knight and Miles Brundage and Mira Murati and Katie Mayer and Peter Welinder and Bob McGrew and Dario Amodei and Sam McCandlish and Ilya Sutskever and Wojciech Zaremba},
      year={2021},
      eprint={2107.03374},
      archivePrefix={arXiv},
      primaryClass={cs.LG},
      url={https://arxiv.org/abs/2107.03374}, 
}

@inproceedings{rodriguez2021evaluation,
    title = "Evaluation Examples are not Equally Informative: How should that change {NLP} Leaderboards?",
    author = "Rodriguez, Pedro  and
      Barrow, Joe  and
      Hoyle, Alexander  and
      Lalor, John P.  and
      Jia, Robin  and
      Boyd-Graber, Jordan",
    editor = "Zong, Chengqing  and
      Xia, Fei  and
      Li, Wenjie  and
      Navigli, Roberto",
    booktitle = "Proceedings of the 59th Annual Meeting of the Association for Computational Linguistics and the 11th International Joint Conference on Natural Language Processing (Volume 1: Long Papers)",
    month = aug,
    year = "2021",
    address = "Online",
    publisher = "Association for Computational Linguistics",
    url = "https://aclanthology.org/2021.acl-long.346/",
    doi = "10.18653/v1/2021.acl-long.346",
    pages = "4486--4503"
}

\clearpage
\appendix
\section{Evaluation Details}
\label{app:eval-details}

Each agent pairs a model with the harness listed in Table~\ref{tab:release-results}, runs at high reasoning effort, and makes four independent attempts per task with a one-hour limit per attempt. An attempt that reaches the time limit is graded on the repository state at that point. For each agent and task, we score the first four graded attempts. Each attempt runs in an isolated container whose only network access is the model provider's API. Cost is the API cost of an attempt as reported by the harness (Claude Code, Codex, and Qwen 3.8 Max), or, where the harness does not report it, computed from the attempt's token usage at the provider's list prices, with the discount for cached input tokens.

\section{Agentic Judge Panel}
\label{app:judge-protocol}

We used panels of agentic judges to curate the programmatic verifiers; the panels do not contribute to the reported scores. Each judge receives the task instruction, a description of the injected vulnerability and of what a correct fix must do, and the reference patch. Each judge then works independently in its own copy of a patched repository with shell access: it can inspect the code, build the project, run its tests, and run the programmatic verifier, whose result is available as evidence. A judge assesses each criterion in Table~\ref{tab:judge-criteria} separately and accepts the patch only if all five hold.

\begin{table}[htb]
\centering
\small
\begin{tabularx}{\linewidth}{lX}
\toprule
Criterion & Requirement \\
\midrule
Security & The repair closes the target vulnerability. \\
Functionality & Ordinary, legitimate behavior remains available. \\
Generalization & The repair blocks attack variants beyond the particular verifier probe. \\
Integrity & The submission repairs the software rather than manipulating the grading mechanism. \\
Soundness & The patched codebase builds and runs in the task environment. \\
\bottomrule
\end{tabularx}
\caption{Judge-panel criteria. A judge accepts a repair only if all five criteria are satisfied.}
\label{tab:judge-criteria}
\end{table}

\section{Human Review}
\label{app:human-review}

During verifier curation, we manually reviewed attempts on which the programmatic verifier and the panel disagreed. The programmatic verifier encodes our definition of a correct repair, so each disagreement was resolved in its favor unless review showed an error in the verifier. When judges showed that the verifier was too strict, failing valid repairs, we replaced the affected checks with checks of behavior. When judges showed that it accepted incomplete repairs, we added probes to catch them. When a judge rejected a patch over an issue that does not affect security or functionality, we kept the verifier's verdict. In addition, every task was reviewed by hand against the checks in Table~\ref{tab:task-checks}.

\section{Weakness Classes and Task Coverage}
\label{app:task-coverage}

The OWASP Top 10:2025 assigns 248 CWEs to its ten categories, but 71 of the 185 CWEs in CWE-bench (38\%) are not assigned to any of them, including most memory-safety weaknesses such as use after free (CWE-416), out-of-bounds write (CWE-787), and type confusion (CWE-843). We therefore assign every CWE in the set to one of ten weakness classes of our own, W01 to W10, defined in the spirit of OWASP but split more finely where the set requires it, for example separating memory and resource safety (W03) and input validation and parsing (W02) from injection (W01). Every task targets a single CWE, and every CWE in the set belongs to exactly one class. The set still covers all ten OWASP Top 10:2025 categories, each with at least two CWEs. Table~\ref{tab:cwe-coverage} lists the CWEs in the evaluation set under their class.

\begin{table}[htb]
\centering
\small
\begin{tabularx}{\linewidth}{l>{\raggedright\arraybackslash}p{2.6cm}rrr>{\raggedright\arraybackslash}X}
\toprule
 & Class & Tasks & Public & CWEs & CWE IDs \\
\midrule
W01 & Injection & 27 & 10 & 20 & 74, 79, 80, 87, 88, 89, 91, 93, 94, 95, 113, 116, 564, 643, 644, 652, 838, 917, 1236, 1427 \\
W02 & Input validation \& parsing & 25 & 4 & 18 & 20, 155, 176, 178, 179, 180, 184, 185, 187, 235, 462, 611, 777, 827, 1286, 1287, 1288, 1289 \\
W03 & Memory \& resource safety & 40 & 15 & 33 & 125, 129, 130, 131, 170, 190, 191, 192, 195, 196, 362, 369, 401, 404, 412, 416, 457, 476, 606, 609, 680, 763, 772, 776, 787, 789, 823, 833, 834, 843, 908, 909, 1333 \\
W04 & Access control & 29 & 10 & 21 & 22, 59, 73, 250, 272, 273, 276, 283, 352, 377, 403, 425, 552, 639, 641, 648, 698, 732, 862, 863, 915 \\
W05 & Authentication \& sessions & 23 & 13 & 18 & 256, 257, 287, 289, 290, 291, 303, 306, 307, 384, 488, 522, 613, 614, 620, 640, 836, 1004 \\
W06 & Cryptography \& secrets & 28 & 16 & 23 & 295, 296, 297, 298, 299, 311, 312, 315, 319, 321, 323, 324, 326, 327, 328, 329, 330, 337, 347, 757, 760, 780, 916 \\
W07 & Trust boundaries \& request handling & 18 & 9 & 14 & 345, 346, 348, 350, 444, 501, 565, 601, 784, 807, 918, 923, 942, 1021 \\
W08 & Supply chain \& data integrity & 18 & 8 & 13 & 426, 427, 434, 454, 471, 472, 494, 502, 829, 1104, 1188, 1357, 1395 \\
W09 & Information exposure \& logging & 25 & 7 & 14 & 117, 200, 201, 209, 212, 214, 219, 226, 359, 433, 524, 532, 598, 1230 \\
W10 & Error handling \& logic flaws & 17 & 8 & 11 & 248, 252, 358, 460, 617, 636, 693, 703, 755, 837, 841 \\\midrule
 & Total & 250 & 100 & 185 & \\
\bottomrule
\end{tabularx}
\caption{CWEs in the evaluation set by weakness class, with the number of tasks, of public tasks, and of distinct CWEs in each class.}
\label{tab:cwe-coverage}
\end{table}

\paragraph{Coverage of the full CWE list.} To show where the classes end, we place every active weakness in CWE v5.0 (956 entries, released October 2026)\footnote{\url{https://cwe.mitre.org/data/downloads.html}} into one of the ten classes or into an out-of-scope group (Table~\ref{tab:cwe-landscape}). A weakness inherits the class of the nearest CWE-bench CWE in the CWE research hierarchy (CWE-1000) when that class is unambiguous, and we assign the remaining entries by hand. Of the 956 weaknesses, 694 (73\%) fall into the ten classes and 262 (27\%) fall outside them. CWE-bench contains 183 of the 694,\footnote{CWE-377 and CWE-613, both in CWE-bench, are deprecated in CWE v5.0 and are not counted.} and its coverage of a class ranges from 16\% for access control (W04) to 50\% for injection (W01).

\paragraph{Out of scope.} The 262 weaknesses outside the classes form six groups. Code-quality and maintainability weaknesses (115), such as dead code, excessive complexity, and missing documentation, do not by themselves make a repository exploitable, so no exploit can separate a vulnerable repository from a repaired one. Hardware-design weaknesses (102), the members of the CWE Hardware Design view, concern circuits, on-chip fabrics, and debug logic rather than software repositories. Intentionally malicious code (8), such as trojan horses and logic bombs, is planted rather than introduced by mistake. User-interface weaknesses (9), such as missing warnings and homoglyphs, depend on how a person perceives the interface, and two weaknesses concern the behavior of machine-learning models rather than the code that calls them. The remaining 26 entries, the eight CWE pillars and broad design classes such as violation of secure design principles (CWE-657), are too general to place in one class; their concrete descendants are classified individually.

\paragraph{Expanding coverage.} CWE-bench can grow in two directions. Within the ten classes, 511 weaknesses have no task yet, most of them in access control (104), memory and resource safety (102), and input validation and parsing (80). Each fits the existing task format of an injected vulnerability checked by an exploit and functionality tests. Beyond the classes, hardware-design weaknesses fit the same format on hardware-description repositories, with simulation testbenches in place of program exploits, and planted malicious code fits an audit task in which the agent must find and remove the code. Code-quality, user-interface, and model-behavior weaknesses have no exploit that confirms a repair, and remain outside the task format.

\begin{table}[htb]
\centering
\small
\begin{tabular}{llrrr}
\toprule
 & Group & CWE v5.0 & CWE-bench & Coverage \\
\midrule
W01 & Injection & 40 & 20 & 50\% \\
W02 & Input validation \& parsing & 98 & 18 & 18\% \\
W03 & Memory \& resource safety & 135 & 33 & 24\% \\
W04 & Access control & 124 & 20 & 16\% \\
W05 & Authentication \& sessions & 55 & 17 & 31\% \\
W06 & Cryptography \& secrets & 56 & 23 & 41\% \\
W07 & Trust boundaries \& request handling & 43 & 14 & 33\% \\
W08 & Supply chain \& data integrity & 29 & 13 & 45\% \\
W09 & Information exposure \& logging & 60 & 14 & 23\% \\
W10 & Error handling \& logic flaws & 54 & 11 & 20\% \\
\midrule
 & In the ten classes & 694 & 183 & 26\% \\
\midrule
-- & Code quality \& maintainability & 115 & -- & -- \\
-- & Hardware design & 102 & -- & -- \\
-- & Too general to classify & 26 & -- & -- \\
-- & Intentionally malicious code & 8 & -- & -- \\
-- & User interface \& human factors & 9 & -- & -- \\
-- & Machine-learning model behavior & 2 & -- & -- \\
\midrule
 & Outside the classes & 262 & -- & -- \\
 & All active weaknesses & 956 & 183 & 19\% \\
\bottomrule
\end{tabular}
\caption{Active weaknesses in CWE v5.0 by weakness class or out-of-scope group, with the number in CWE-bench and the share of each class it covers.}
\label{tab:cwe-landscape}
\end{table}

\section{Released Artifacts}
\label{app:release}

The 100 public tasks are released as a Hugging Face dataset (\url{https://huggingface.co/datasets/collinear-ai/cwe-bench}). Each task is a self-contained Harbor task comprising a Dockerized environment with the repository, the audit instruction, the deterministic verifier, and the reference patch. The dataset ships a registry, so the full public set can be run directly from the Hub:
\begin{verbatim}
harbor run --repo "https://huggingface.co/datasets/collinear-ai/cwe-bench" \
  -d cwe-bench@2.0 -k 4 --agent <agent> --model <model> --env docker
\end{verbatim}
We also release the 4{,}400 rollouts behind the public-set results in Table~\ref{tab:release-results} (eleven agents, four attempts per task) at \url{https://huggingface.co/buckets/collinear-ai/cwe-bench-rollouts}, with their trajectories, final patches, and verifier logs, an index of every rollout with its reward, token counts, and cost, and a script that recomputes the public-set pass@1, pass@4, and cost in Table~\ref{tab:release-results}. A leaderboard is maintained at \url{https://cwe-bench.com} and \url{https://huggingface.co/spaces/collinear-ai/cwe-bench-leaderboard}. The 150 held-out tasks are not released; agents can be evaluated on them on request.

\section{Item Response Theory Analysis}
\label{app:irt-artifacts}

As another view of the results, we fit a Rasch item response theory (IRT) model~\citep{rodriguez2021evaluation} to the outcomes on all 250 tasks in Table~\ref{tab:release-results}. The model treats the passing attempts $c_{mj}$ of agent $m$ on task $j$ out of $n_{mj}$ as
\begin{equation}
c_{mj}\sim\operatorname{Binomial}(n_{mj},p_{mj}),\qquad p_{mj}=\frac{1}{1+\exp\{-(\theta_m-b_j)\}},
\label{eq:irt}
\end{equation}
where $\theta_m$ is the ability of agent $m$ and $b_j$ the difficulty of task $j$. Abilities are centered at zero, and intervals come from 300 bootstrap refits.

\begin{figure}[!htbp]
\centering
\includegraphics[width=\linewidth]{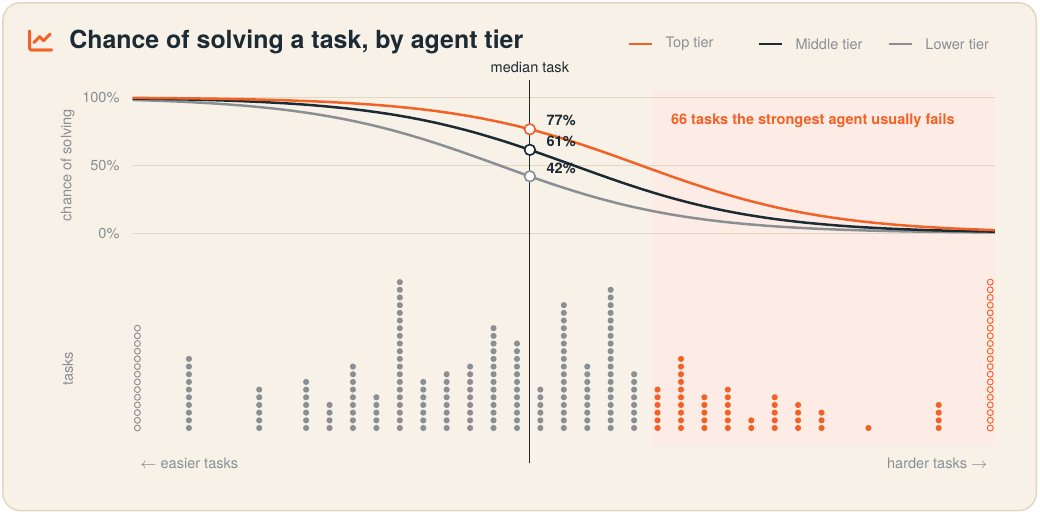}
\caption{Predicted chance of solving a task as a function of task difficulty, from a Rasch model fit to the outcomes on all 250 tasks in Table~\ref{tab:release-results}. Each curve uses the mean ability of one tier: top (Claude Opus 5.5, Muse Wonder, Gemini 4 Argon, GPT-6 Astra), middle (DeepSeek V4.1 Flash, Grok 4.7), and lower (Hy4 Preview, Mistral Large 4, Qwen 3.8 Max, Kimi K3.1 Pro, GLM 5.3). The vertical line marks the task of median difficulty. Below, each circle is one task placed by its estimated difficulty; hollow circles are tasks that every agent always or never solved. The shaded region contains the 66 tasks the strongest agent is predicted to solve less than half the time.}
\label{fig:irt-rasch}
\end{figure}

Because every agent attempted every task four times, the fitted abilities follow the pass@1 ordering; what the model adds is a placement of tasks and agents on the same scale. Figure~\ref{fig:irt-rasch} shows the fitted success curves for three tiers of agents. Claude Opus 5.5, Muse Wonder, Gemini 4 Argon, and GPT-6 Astra have overlapping intervals, above a middle tier of DeepSeek V4.1 Flash and Grok 4.7 and a lower tier of the remaining five agents. On the task of median difficulty, the predicted chance of success is 77\%, 61\%, and 42\% for the three tiers. Sixty-six tasks have an estimated difficulty above the strongest agent's ability (48 to 78 across bootstrap refits, 95\% interval), meaning even the strongest agent is predicted to solve them less than half the time.

\end{document}